# DETECTION AND QUANTIFICATION OF VOLATILES AT MARS USING A MULTISPECTRAL LIDAR

A. J. Brown[1], T. Michaels[1], L. Fenton[1]

[1]SETI Institute, Mountain View, CA (corresponding author abrown@seti.org),

P.O. Hayne[2], S. Piqueux[2]

[2]Jet Propulsion Laboratory, California Institute of Technology

T.N. Titus[3]

[3]USGS Flagstaff, AZ

M.J. Wolff[4], R. T. Clancy[4], G. Videen[4]

[4]SSI, Boulder, CO

W. Sun[5]

[5]NASA Langley

R. Haberle[6], A. Colaprete[6]

[6]NASA Ames Research Center

M.I. Richardson[7]

[7]Aeolis Research, Los Angeles, CA

S. Byrne[8]

[8]LPL Univ of Arizona

R. Dissly[9]

[9]Ball Aerospace, CO

S. Beck[10]

[10]Aerospace Corp, CA

C. Grund[11]

[11]Lightworks, LLC, CO

**Introduction**

We present a concept for using a polarization sensitive multispectral lidar such as the *ASPEN* instrument proposed in [1] to map the seasonal distribution and exchange of volatiles among the reservoirs of the Martian surface and atmosphere.

**Concept**

The *ASPEN* instrument will be a multi-wavelength, altitude-resolved, active near-infrared (NIR, with 10 bands around 1.6 microns) instrument to measure the reflected intensity and polarization of backscattered radiation from planetary surfaces and atmospheres. The proposed instrument would be ideally suited for a mission to Mars to comprehensively investigate the nature and seasonal distributions of volatiles and aerosols. The investigation would include the abundance of atmospheric dust and condensed volatiles, surface and cloud/aerosol grain sizes and shapes, ice and dust particle microphysics and also variations in atmospheric chemistry during multiple overflight local times throughout polar night and day.

Such an instrument would be ideal for mapping and detection of frost phenomena [2] and precipitation events [3] in the polar regions of Mars. Herein we discuss the applicability of this instrument to detect sublimation/deposition 'mode flips' reported in [5]. A full range of scientific questions to be addressed by this instrument is presented in [1].

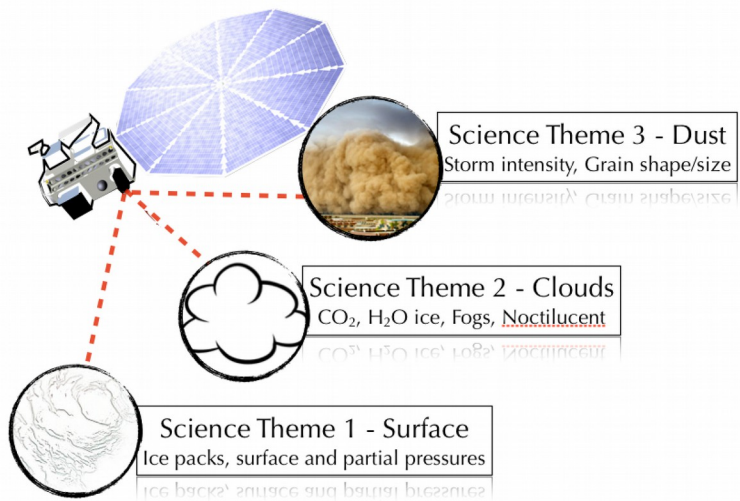

**Figure 1 – Multispectral lidar concept in orbit around Mars, with science themes of Surface, Clouds and Dust**

**Cubesat opportunity**

Although the full scale multispectral lidar requires a 1m receiver mirror that dictates space and weight of the instrument by today's technological standards, an opportunity exists to carry out a pathfinder mission with a cubesat footprint similar to that used on the Lunar Flashlight mission [4]. Lunar Flashlight utilizes a multi-band laser reflectometer to measure the surface reflectance, thereby demonstrating this multiband lidar concept on a small spacecraft in lunar orbit. If payload space becomes available in the coming decade for Martian cubesat class missions, for example as part of a SpaceX ridealong mission, we would like to exploit this for a trispectral lidar (at least 3 bands) and perform a proof of the concept of the *ASPEN* mission that provides some of the science discussed here (e.g. high altitude $H_2O$ clouds and lower spatial resolution surface $H_2O$ ice) for a reduced cost.

**Previous work with passive hyperspectral instrument**

As reported in [5], we have used observations from the Compact Reconnaissance Imaging Spectrometer for Mars (CRISM) of the north polar cap during late summer for four Martian years, to monitor the summertime water cycle in order to place quantitative limits on the amount of water ice deposited and sublimed in summer. The most compelling result of this map is that we have identified regions and periods of 'net deposition' and 'net sublimation' on the summer north polar cap. Regions of the cap undergo a 'mode flip' from sublimation to deposition mode and the timing of ***mode flips*** is latitude dependent. This enables us to place firmer estimates on the dynamics by using the concept of depositional mode flips, a previously unknown observable that is also applicable to testing and verifying Martian Global Climate Models (GCMs).

**$H_2O$ index volatile tracking**

Previous work has tracked the variations in the so called $H_2O$ index [5-9] over the parts of the cap that received CRISM coverage throughout the summer period over four Mars Years. The index is based on the depth of the water ice 1.5μm absorption band. It is high when water ice is present, and grows with the water ice grain size. When deposition of fine grained ice occurs, the $H_2O$ index decreases, because finer grained ice scatters light back to the observer more readily and in turn decreases the depth of the 1.5μm $H_2O$ absorption band [10].

**Applicability of a multispectral lidar**

As described in detail in [1], a 10 band NIR multispectral lidar system can carry out the same measurements of atmospheric volatiles as CRISM in the polar regions, and is in fact more sensitive when the multispectral bands are chosen effectively. Not only will the lidar produce finer maps of the $H_2O$ index (and a $CO_2$ index), but those indexes can be extended into the polar nighttime, thus extending our knowledge of the distribution of polar volatiles throughout the year. Finally, the lidar will provide time resolved measurements, allowing discrimination of clouds and fog, a task which is very difficult for CRISM and other passive instruments. As with the MOLA instrument, surface elevation can be measured to determine seasonal cap thicknesses and mass wasting processes on longer timescales.

**Previous work on brightening of north polar cap**

A long-standing problem of the Martian climate is the summer brightening of the north polar cap. This was first reported by Kieffer [11] using IRTM data, and subsequently observed with TES by Kieffer and Titus [12]. Bass and Paige [13] used IRTM and MAWD measurements to determine the peak of water vapor over the north polar cap. They found that the lowest visible albedo occurred during $L_s$=93-103° and water vapor was also released after $L_s$=103°; however they could not determine whether this was caused by changes in water ice grain size or dust deposition.

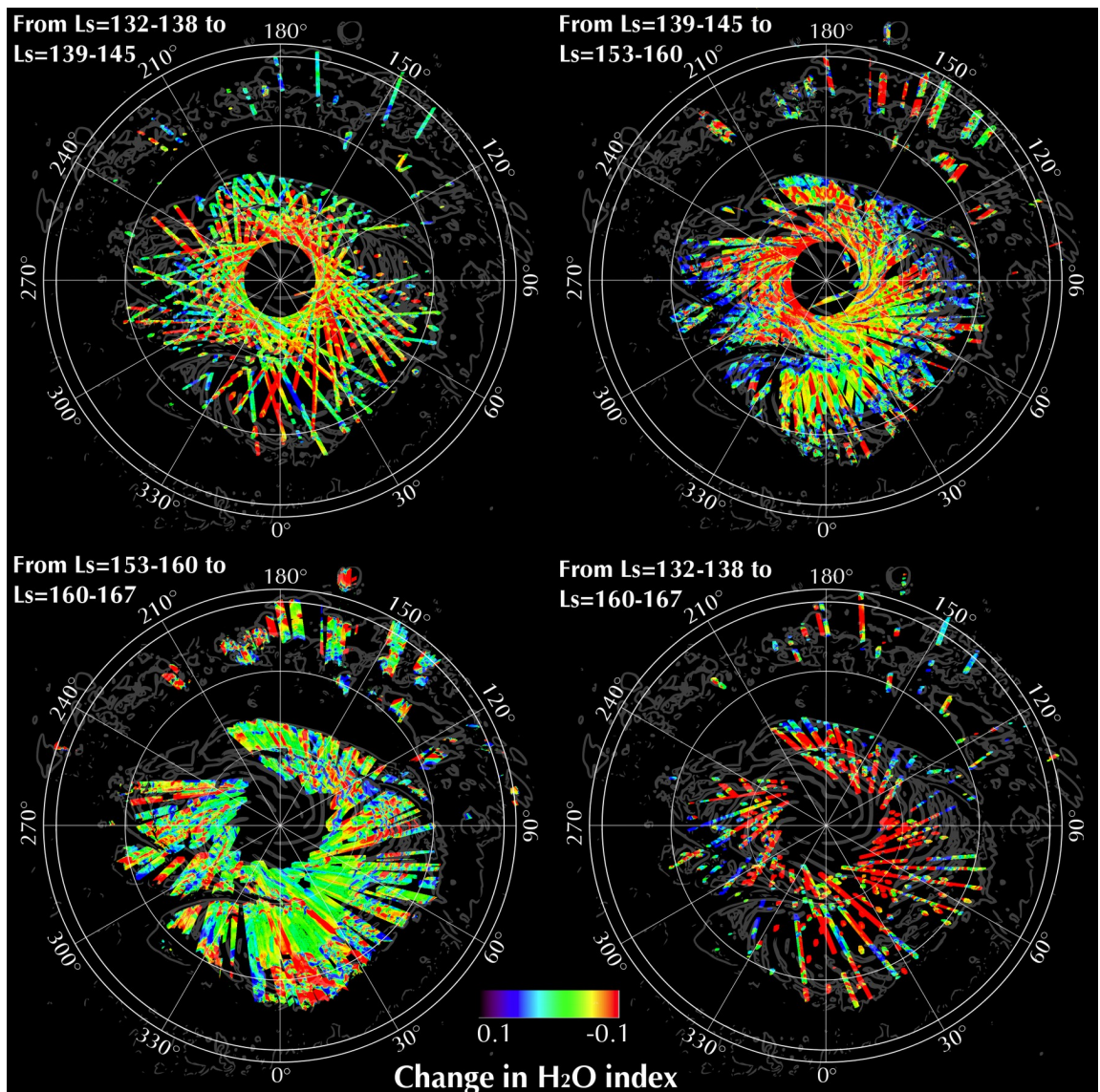

**Figure 2** – Changes in H$_2$O index for MY28 showing net deposition in red colors and net sublimation in blue. The bottom right image is a summary of the whole period from L$_s$=132 to L$_s$=167. From [5].

**Deposition/Sublimation 'Mode flips'**

We previously used CRISM H$_2$O index maps (Figure 2) to show that in a key region in the interior of the north polar cap, the absorption band depths grow until L$_s$=130°, as reported in [5], followed by a period when they begin to shrink, until they are obscured at the end of summer by the north polar hood (Figure 2). This behavior is transferable over the entire north polar cap, where in late summer regions 'flip' from being net sublimating into net condensation mode as the weather cools (Figure 3). This 'mode flip' happens earlier for regions closer to the pole, and later for regions close to the periphery of the cap. For some parts of the periphery of the cap, there are regions where water ice absorption band depths have not been observed to decrease over the time we have observed them, suggesting that they may remain in net sublimation mode during the entire summer season and only go into condensation mode in winter.

**Total deposition of water ice during summer**

Under the assumption that the observed shrinking of grain sizes is entirely due to the deposition of fine grained water ice, we have approximated the total amount of water ice deposited on the cap each summer, which equates to 70 microns of deposition over the $L_s$=132-168° late summer period. This amount is considerably more than the ~6 microns of deposition of water ice on the south polar cap during the summer period as reported in [14,15].

**Conclusions**

A multispectral lidar could make fundamentally new observations of the Martian surface and atmosphere to quantify the deposition of volatiles throughout the entire Martian year at an unprecedented spatial resolution. We have briefly introduced the water absorption band maps made using CRISM for the entire north polar region as a function of space and time over late summer which identified 'net deposition' and 'net condensation' regions and periods. This provides a tantalizing glimpse into what a multispectral lidar in orbit around Mars would reveal that would be crucial to understanding the long term Martian volatile inventory and dynamics.

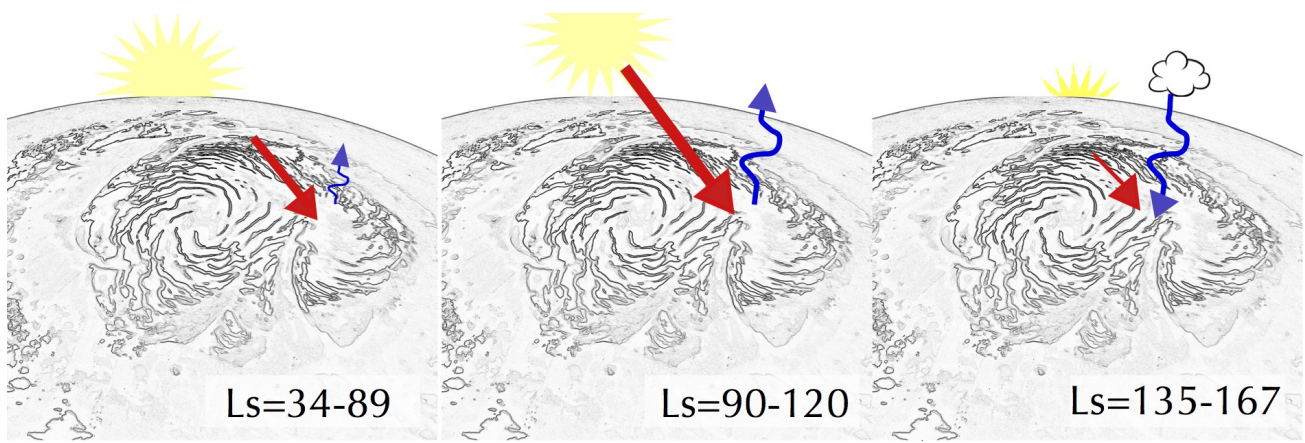

**Figure 3** – Cartoon representation of deposition/sublimation 'mode flips'. Dates given are relevant for the Gemini Lingula region (where the arrows point in the right image).

**Take home messages**

**1.** Previous studies have identified regions and periods of net deposition and net sublimation on the Martian polar caps [5]. We have identified an instrumental concept that could be flown to Mars in order to expand our understanding of the dynamics of Martian volatiltes. Opportunities exist for a precursor active LIDAR instrument to flown as a cubesat.

**2.** Studies such as [1, 6-8] have revealed the path forward for investigations into the transport of water in the Martian climate cycle. Using CRISM observations, we have now quantified the spring and summer water ice deposition for both poles. These measurements are crucial to our understanding of the construction and ongoing stability of the caps under today's climate. However, there is a ***clear and pressing*** need to understand the fall and winter 'dark side' of the Martian polar region that is impenetrable to passive instruments like CRISM and MARCI and instead requires multi-wavelength lidar instruments such as the *ASPEN* concept discussed here.

**Acknowledgements**